\documentclass[journal]{IEEEtran}

\ifCLASSINFOpdf
\else
\usepackage[dvips]{graphicx}
\fi
\usepackage{url}

\usepackage{tabularx}
\usepackage{multirow}

\newcommand{\mytable}{
	\centering
	\renewcommand{\arraystretch}{1.2}
}
\newcommand{\captionsep}{\vspace*{-5pt}}
\newcolumntype{L}{>{\raggedright\arraybackslash}X}
\newcolumntype{C}{>{\centering\arraybackslash}X}
\usepackage{booktabs}

\usepackage{graphicx}

\usepackage{amsmath}

\usepackage{hyperref}

\usepackage{cite}

\usepackage{microtype}

\usepackage{xcolor}
\definecolor{hermancolor}{HTML}{FF6600}

\begin{document}
	
	\title{Leveraging Multilingual Transfer for Unsupervised Semantic Acoustic Word Embeddings}
	
	\author{Christiaan Jacobs, \IEEEmembership{Student Member, IEEE}, Herman Kamper, \IEEEmembership{Senior Member, IEEE}
		
		\thanks{Manuscript submitted to IEEE Signal Processing Letters.}
		\thanks{The authors are with E\&E Engineering, Stellenbosch University, South Africa (e-mail: 20111703@sun.ac.za; kamperh@sun.ac.za).}}
	
	\markboth{Submitted 2023}
	{Shell \MakeLowercase{\textit{et al.}}: Bare Demo of IEEEtran.cls for IEEE Journals}
	
	\maketitle
	
	\begin{abstract}
		Acoustic word embeddings (AWEs) are fixed-dimensional vector representations of speech segments that encode phonetic content so that different realisations of the same word have similar embeddings.
		In this paper we explore semantic AWE modelling.
		These AWEs should not only capture phonetics but also the meaning of a word (similar to textual word embeddings).
		We consider the scenario where we only have untranscribed speech in a target language.
		We introduce a number of strategies leveraging a pre-trained multilingual AWE model---a phonetic AWE model trained on labelled data from multiple languages excluding the target.
		Our best semantic AWE approach involves clustering word segments using the multilingual AWE model, deriving soft pseudo-word labels from the cluster centroids, and then training a Skipgram-like model on the soft vectors.
		In an intrinsic word similarity task measuring semantics, this multilingual transfer approach outperforms all previous semantic AWE methods.
		We also show---for the first time---that AWEs can be used for downstream semantic query-by-example search.
	\end{abstract}
	
	\begin{IEEEkeywords}
		Semantic embeddings, acoustic word embeddings, semantic retrieval, query-by-example search.
	\end{IEEEkeywords}
	
	\section{Introduction}

\IEEEPARstart{W}{ord} embedding models such as Word2Vec~\cite{mikolov_efficient_2013, mikolov_distributed_2013} and GloVe~\cite{pennington_glove_2014} revolutionised natural language processing (NLP) by mapping written words to continuous fixed-dimensional vectors.
These models learn from co-occurrence information in large unlabelled text corpora.
As a result, words that are related in meaning end up having similar embeddings.
This has led to improvements in a wide range of NLP tasks~\cite{sebastiani_machine_2002, manning_introduction_2008, mikolov_exploiting_2013, conneau_word_2018}. 
However, limited efforts have been made to generate such semantic representations for spoken words.

While acoustic word embedding (AWE) models~\cite{levin_fixed-dimensional_2013, settle_discriminative_2016} map variable-duration speech segments to fixed-dimensional vectors, these models do not aim to capture meaning.
The goal is rather to map different realisations of the same word to similar embeddings, i.e.\ the embedding space encodes phonetic rather than semantic similarity.
Several unsupervised AWE modelling techniques have been explored~\cite{chung_unsupervised_2016, kamper_deep_2016, settle_query-by-example_2017, kamper_truly_2019, jacobs_acoustic_2021}.
Recently, multilingual AWE models have been introduced as an alternative~\cite{kamper_multilingual_2020, hu_multilingual_2020,kamper_improved_2021, jacobs_multilingual_2021}: a single AWE model is trained on labelled data from multiple well-resourced languages and then applied to an unseen target low-resource language.

\begin{figure}[!t]
\centerline{\includegraphics[width=.85\columnwidth]{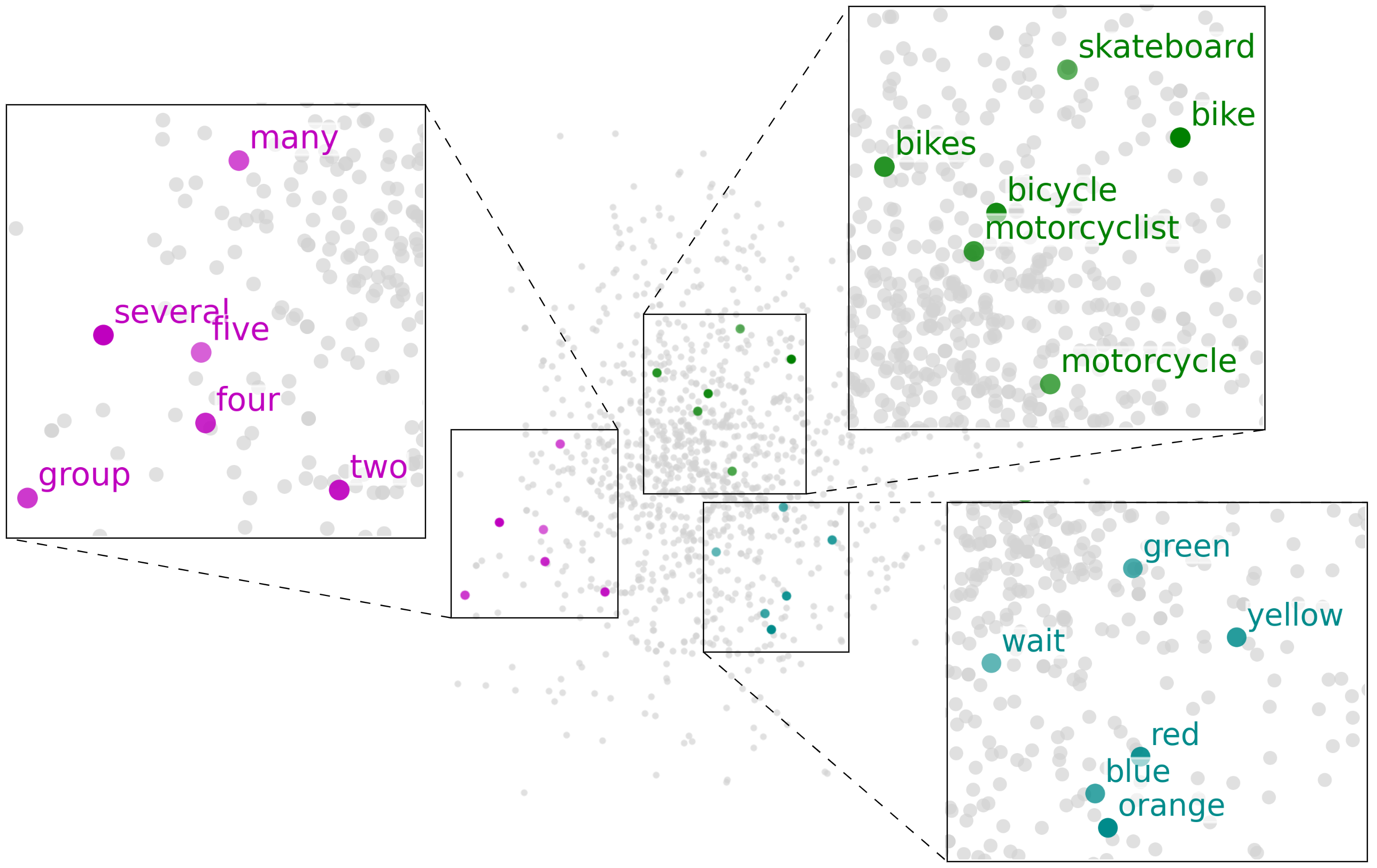}}
\caption{
PCA projection of semantic AWEs (averaged), produced by our Cluster+Skipgram model on development data.
We highlight the five nearest neighbours for ``several" (pink), ``bike" (green), and ``orange" (blue). 
}
\label{fig:sem_space}
\end{figure}

While phonetic AWEs have proven useful in several downstream applications~\cite{kamper_embedded_2017,jacobs_towards_2023}, there are also many cases where semantics would be beneficial.
In semantic AWE modelling the goal would be to map speech segments to vector representations that not only capture whether two segments are instances of the same word, but also the semantic relationship between words. 
E.g., we want an AWE space where different instances of ``red'' are close to each other, but also close to instances of ``blue". 
And all these embeddings should be far from unrelated words, such as ``group''.
An example is given in Fig.~\ref{fig:sem_space}, which visualises actual AWEs from our approach.

Learning semantic AWEs from speech is challenging due to channel variability, noise, and speaker-specific information that are not present in written text.
Some studies, therefore, use another modality as a grounding signal, 
e.g.\ using images~\cite{harwath_deep_2015, kamper_visually_2017, kamper_semantic_2019} or text labels~\cite{abdullah_integrating_2022} as a weak form of supervision.
Only a handful of studies have looked at learning semantic AWEs from unlabelled speech alone~\cite{chung_speech2vec_2018, chen_phonetic-and-semantic_2019}.
To overcome these challenges, we propose leveraging the recent improvements in multilingual modelling for phonetic AWEs.

We specifically propose using transfer learning from a phonetic multilingual AWE model to obtain a semantic AWE model in a target language where we only have unlabelled speech.
Since the multilingual model already captures phonetics, this should simplify the semantic learning problem.
We present three approaches.
Our best approach involves using a multilingual AWE model to cluster unlabelled word segments from the target language. 
For each segment, we derive a soft pseudo-word label vector based on the proximity to the cluster centroids.
Finally, we get semantic AWEs by training a Skipgram-like model on these soft vectors.

In an intrinsic word similarity task, this approach outperforms previous methods learning from scratch~\cite{chung_speech2vec_2018, chen_phonetic-and-semantic_2019} and also our other multilingual transfer methods.
We also show that this method can be used downstream in an extrinsic semantic query-by-example search task.

	\label{sec:awe_models}

\section{Phonetic Acoustic Word Embeddings}
\label{ssec:p_awe}
Most existing AWE methods map speech segments to a vector space where instances of the same word class are located near each other.
We call these \textit{phonetic AWEs}, because the space should capture whether input segments are phonetically similar rather than related in meaning.
Formally, a speech segment $X = \left( \mathbf{x}_1, \mathbf{x}_2, \ldots, \mathbf{x}_T \right)$ is projected to a vector $\mathbf{z}$, with each $\mathbf{x}_t$ a speech frame. 
Two phonetic AWE models have proven to be particularly effective: the correspondence autoencoder RNN (CAE-RNN)~\cite{kamper_improved_2021} and the ContrastiveRNN~\cite{jacobs_acoustic_2021}. 

\textbf{CAE-RNN.} This model uses an encoder RNN to map a word segment $X$ to a latent embedding $\mathbf{z}$.
The embedding $\mathbf{z}$ is then given to a decoder RNN to reconstruct a target word segment $X^{\prime}$, where $X^{\prime}$ is a different instance of the same class as the input.
The model is optimised to minimise the reconstruction loss, ${\sum_{t=1}^{T^{\prime}} \lVert \mathbf{x}_t^\prime - \boldsymbol{f}_t(X) \lVert^2}$, where $\boldsymbol{f}_t(X)$ is the $t$th decoder output conditioned on embedding $\mathbf{z}$.
During inference, the encoder generates a unique AWE $\mathbf{z}$ for every new input segment.

\textbf{ContrastiveRNN.} 
This model explicitly minimises the distance between embeddings from speech segments of the same word class while maximising the distance between words of a different class. 
Formally, given speech segments $X_{\text{anc}}$ and $X_{\text{pos}}$ containing instances of the same word class and multiple negative examples $X_{\text{neg}_1},\ldots, X_{\text{neg}_N}$, the ContrastiveRNN produces embeddings $\mathbf{z}_{\text{anc}}, \mathbf{z}_{\text{pos}}, \mathbf{z}_{\text{neg}_1}, \ldots, \mathbf{z}_{\text{neg}_N}$.
The loss is then defined as~\cite{chen_simple_2020}:
\begin{equation}
J = -\text{log}\frac{\text{exp}\left(\text{sim}(\mathbf{z}_{\text{anc}}, \mathbf{z}_{\text{pos}})/\tau\right)}{\sum_{j \in \{\text{pos}, \text{neg}_1, \hdots, \text{neg}_N\}}^{}\text{exp}\left(\text{sim}(\mathbf{z}_{\text{anc}}, \mathbf{z}_j)/\tau\right)}
\label{eqn:contrastive_loss}
\end{equation} 
where $\text{sim}(\cdot)$ denotes cosine similarity and $\tau$ is a temperature parameter, tuned on development data.

From this point onwards we add the subscript \text{p} to the AWEs described in this section, i.e.\ $\mathbf{z}_{\text{p}}$ indicates that the embedding preserves phonetic information related to word class only.

Previous studies showed the advantage of using these models in a multilingual transfer setup where a single AWE model is trained on labelled data from multiple well-resourced languages before transferring and applying it to an unseen target language~\cite{kamper_improved_2021, jacobs_acoustic_2021}.
This allows for AWEs to be obtained even in languages for which we do not have any labelled data.

\section{Semantic AWEs (Trained from Scratch)}
\label{ssec:s_awe}

Two approaches have been proposed that adapt the framework above to obtain semantic embeddings $\mathbf{z}_{\text{s}}$, where the embeddings not only reflect phonetic similarity but also capture meaning.
In both cases~\cite{chung_speech2vec_2018, chen_phonetic-and-semantic_2019}, the 
problem is simplified by assuming that we know where words start and end
(but the word classes are still unknown), i.e. we have an unlabelled speech corpus $\{X^{(n)}\}^{N}_{n=1}$ of $N$ segmented word tokens. We also make this assumption.

\textbf{Speech2Vec}~\cite{chung_speech2vec_2018} is a variant of the CAE-RNN
where, instead of using pairs of instances of the same word class, the positive pairs are now context word pairs $(X_{\text{trg}}, X_{\text{ctx}})$.
$X_{\text{trg}}$ is a target centre word segment while $X_{\text{ctx}}$ is a context word appearing somewhere in a window around the centre.
These context pairs  are constructed without word labels by only considering the relative position of words within an utterance.
Speech2Vec was inspired by the Skipgram model for text data~\cite{mikolov_efficient_2013}, where an input word is fed to a log-linear classifier that predicts words within a context window.
By using the CAE-RNN reconstruction loss, Speech2Vec similarly tries to reconstruct a word segment that appears near its input. 
Ideally the resulting embeddings $\mathbf{z}_{\text{s}}$ should therefore be similar for words that co-occur in the speech data.
There have recently been concerns about the original Speech2Vec implementation~\cite{chen_reality_2023}, and we therefore use our own version here (but still refer to it as Speech2Vec).

By similarly modifying the model presented in Sec.~\ref{ssec:p_awe}, a \textbf{semantic ContrastiveRNN} can be trained on target (anchor), context (positive), and out-of-context word segments (negatives) to learn semantic embeddings using the loss in~\eqref{eqn:contrastive_loss}.
This approach is similar to~\cite{chen_phonetic-and-semantic_2019}, where they include a trainable network to remove speaker information.

In both these methods, a semantic AWE model is trained from scratch, therefore requiring the models to learn to capture phonetic and semantic similarity simultaneously.

\section{Our Approach: Using Multilingual Transfer for Semantic AWEs}
\label{sec:multi}

Our new proposal is to utilise a pre-trained multilingual AWE model (end of Sec.~\ref{ssec:p_awe}) to assist semantic AWE modelling.
Three specific strategies are proposed.

\textbf{ContrastiveRNN with multilingual initialisation.}
Instead of training semantic models from scratch~(\ref{ssec:s_awe}), we can warm-start them using the learned weights of a pre-trained multilingual AWE model.
In our experiments, we use the learned weights of a multilingual AWE model's encoder to initialise the encoder RNN of the ContrastiveRNN.
The model is then updated on context pairs from the target language using \eqref{eqn:contrastive_loss}. 

\textbf{Projecting multilingual AWEs.} Alternatively, we can project an existing phonetic AWE space to a new semantic AWE space.
First, we apply the multilingual model to the unlabelled speech segments $\{X^{(n)}\}$ to get a set of phonetic AWEs $\{\mathbf{z}^{(n)}_{\text{p}}\}$.
Then we train a projection network that maps the phonetic AWEs to semantic embeddings $\{\mathbf{z}^{(n)}_{\text{s}}\}$.
The projection network is trained using the contrastive loss \eqref{eqn:contrastive_loss}, optimising the distances between the output embeddings $\mathbf{z}_{\text{s}}$.

\begin{figure}[!t]
\centerline{\includegraphics[width=0.9\columnwidth]{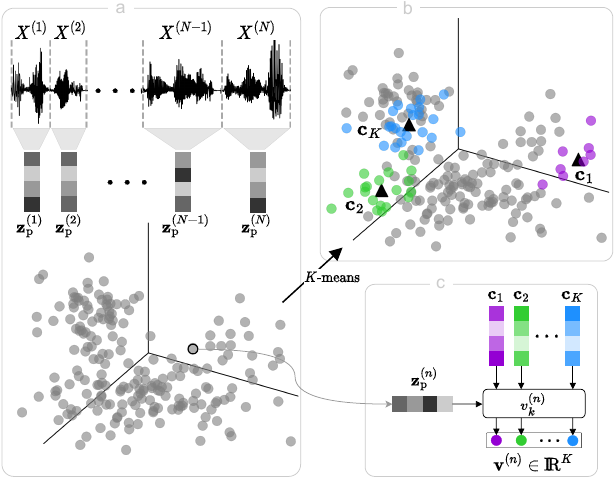}}
\caption{
    Our Cluster+Skipgram semantic AWE approach.
    Speech segments $X^{(n)}$ are represented by soft pseudo-word label vectors $\mathbf{v}^{(n)}$ which are then used to train a Skipgram-like model.
    }
\label{fig:cluster}
\end{figure}

\textbf{Cluster+Skipgram.}
This approach is based on the Skipgram Word2Vec model~\cite{mikolov_efficient_2013}.
Instead of using a fixed dictionary of discrete word class labels to construct input and output vectors to train a Skipgram model on text, we use the phonetic similarities in the original AWE space to derive a soft pseudo-word label for each speech segment.
This is illustrated in Fig.~\ref{fig:cluster}.
In more detail, a multilingual AWE model is applied to the segmented speech corpus $\{X^{(n)}\}$, producing a set of phonetic AWEs $\{\mathbf{z}_\text{p}^{(n)}\}$.
Next we apply $K$-means clustering to the phonetic embedding space, producing a set of centroids $\{\mathbf{c}_{k}\}^{K}_{k=1}$.
The idea is that these clusters should resemble distinct word classes.
We then calculate a soft vector label of an AWE belonging to each cluster:
\begin{equation}
v_{k}^{(n)} = \frac{\exp\left(-\text{sim}(\textbf{z}_{\text{p}}^{(n)}, \textbf{c}_k)/\sigma^2\right)}{\sum_{j=1}^{K}\exp\left(-\text{sim}(\textbf{z}_{\text{p}}^{(n)}, \textbf{c}_j)/\sigma^2\right)}
\label{eq:prob_dist}
\end{equation}
where $\text{sim}(\cdot)$ denotes cosine similarity and $\sigma$ is a hyperparameter controlling the influence of distant centroids.
Each segment is represented by a unique vector $\mathbf{v}^{(n)}$, with segments from the same word class ideally having similar representations.
This is different from Word2Vec, where a single one-hot vector represents a unique word class.
Finally, a linear classifier model is trained with these continuous vectors as input and target outputs using the negative log-likelihood loss (as in the original Skipgram).
We also experimented with hard clustering, but this gave very poor performance on development data.

	\section{Experimental Setup}
\label{sec:setup}

\textbf{Data.} 
We perform experiments using the Flickr8k Audio Captions Corpus (FACC)~\cite{harwath_deep_2015}.
This corpus contains 40k spoken captions in English describing the content of a Flickr image~\cite{rashtchian_collecting_2010}.
This is useful for measuring semantics: the images come from a fairly narrow domain, and the semantic concepts, therefore, reoccur in different utterances.
We do not use the images during training: the spoken captions are treated as our unlabelled target speech corpus.
We use the default train, development, and test splits containing 30k, 5k, and 5k spoken utterances, respectively.
Speech audio is parametrised as 13-dimensional static mel-frequency cepstral coefficients (MFCCs).
We also perform experiments using self-supervised speech features:
we use the 12th transformer layer of the multilingual XLSR model~\cite{babu_xls-r_2022} to get 1024-dimensional features.
Previous work has shown that self-supervised speech features (obtained in an unsupervised way) can be useful as the frame-level input to AWE models~\cite{van_staden_comparison_2021, sanabria_analyzing_2023}.
Utterances are normalised per speaker and segmented using true word boundaries from forced alignments~\cite{mcauliffe_montreal_2017}.
We use these word segments to construct context word pairs as described in Sec.~\ref{ssec:s_awe}.
For all the semantic models, we use a context window of three words before and after a centre word.

\textbf{Semantic models trained from scratch (\ref{ssec:s_awe}).}
The encoder and decoder of our Speech2Vec implementation each consist of three unidirectional RNNs with 400-dimensional hidden vectors and an embedding size of 100.
The model is trained on roughly two million word pairs occurring in the same context window in our training data.
The semantic ContrastiveRNN uses the same encoder structure. 
It is also trained on the same context pairs together with out-of-context word segments serving as negatives; for each positive, we sample 20 negatives.

\textbf{Semantic models using multilingual transfer (\ref{sec:multi}).}
We train a CAE-RNN multilingual AWE model~\cite{kamper_improved_2021, jacobs_acoustic_2021} on five different Common Voice~\cite{ardila_common_2020} languages: Italian, Dutch, Russian, Czech.
We pool the data from all languages and extract 300k training pairs. 
The CAE-RNN model structure is the same as that of our Speech2Vec model.
This multilingual CAE-RNN is used to initialise a ContrastiveRNN;
we freeze the weights of the first two encoder layers while training on context pairs.
For the projection network, we use a feed-forward network of two linear layers with an inner dimension of 1024 and input and output dimensions of 100.
Again we sample 20 negatives for each positive and train the network to optimise the contrastive loss~\eqref{eqn:contrastive_loss}.
For the Cluster+Skipgram approach, we use the multilingual CAE-RNN to obtain phonetic embeddings.
For $K$-means clustering, we use $K = 5000$ clusters and set $\sigma = 0.01$ in \eqref{eq:prob_dist}.
We use the same linear network as the Skipgram model~\cite{mikolov_efficient_2013}, with a word embedding size of 100 and optimise the network with the negative log-likelihood loss.

\textbf{Intrinsic evaluation.} We evaluate the quality of semantic embeddings by measuring similarity scores between isolated word pairs.
We compare these scores to word similarity scores of textual word embeddings generated by an off-the-shelf Skipgram model trained on the transcribed utterances. 
Spearman's $\rho$ is used to quantify the similarity between the two sets of word-pair similarities~\cite{agirre_study_2009 ,hill_simlex-999_2015}. 
To obtain a single semantic embedding for each word class, we calculate the average of all AWEs from the same class and report $\rho_{\text{avg}}$. 
Given that we are particularly interested in obtaining semantic embeddings for individual word segments, single-sample performance is also measured by randomly selecting one instance of each spoken word.
This is repeated ten times and averaged to get a single score $\rho_{\text{single}}$.

\textbf{Extrinsic evaluation.} 
We use the setup as \cite{kamper_semantic_2019} to evaluate downstream semantic query-by-example (QbE) search performance.
Semantic labels for 1000 test utterances from FACC were collected from human annotators, using a set 67 keyword classes.
Specifically, each of the 1000 utterances was labelled by five annotators,
indicating whether a particular keyword is semantically relevant to that utterance (regardless of whether the word instance appears verbatim in the utterance).
We use the majority decision to assign a hard label for whether a query keyword is relevant to an utterance.
Using these hard labels, we calculate semantic $P@10$, $P@N$, EER, and Spearman's $\rho$. 
Here, $\rho$ measures the correlation between a system's ranking and the actual number of annotators who deemed a query keyword relevant to an utterance. 
To simplify the QbE task, we still assume that ground truth word boundaries are known: a query AWE is therefore compared to AWEs for the word segments in an unlabelled search utterance.

	\section{Results}

\begin{table}[!t]
    \mytable
    \caption{    
    Spearman's $\rho$ (\%) for semantic word similarity of AWEs on development data. 
    Embeddings are either averaged over word segments of the same class, or a single segment is sampled per class. AWE models take either MFCC or XLSR features as input.
    }
    \captionsep
    \begin{tabularx}{\linewidth}{@{\extracolsep{2pt}}L*{4}{c}}
    \toprule				
        & \multicolumn{2}{c}{MFCC} & \multicolumn{2}{c}{XLSR} \\
        \cmidrule(l){2-3} 
        \cmidrule(l){4-5}
       Model & { $\rho_{\text{single}}$ }&   $\rho_{\text{avg}}$ &  $\rho_{\text{single}}$ &   $\rho_{\text{avg}}$ \\ 
    \midrule

        \multicolumn{2}{@{}l}{\underline{\textit{Trained from scratch~(\ref{ssec:s_awe}):}}} \\ 

        Speech2Vec & \hphantom{0}1.0 & \hphantom{0}5.3 & \hphantom{0}7.2 & 21.3 \\
        ContrastiveRNN & \hphantom{0}0.7 & \hphantom{0}5.1 & \hphantom{0}4.6 & 25.9 \\[1mm]


        \multicolumn{2}{@{}l}{\underline{\textit{Using multilingual transfer~(\ref{sec:multi}):}}} \\
        
        ContrastiveRNN multilingual init. & \hphantom{0}0.6 & \hphantom{0}5.5 & \hphantom{0}6.1  & 24.5 \\
        Project AWE & \hphantom{0}3.6 & 18.2 & 18.3 & 33.6 \\
        Cluster+Skipgram & \textbf{18.0} & \textbf{35.6} & \textbf{35.9} & \textbf{41.7} \\
    \bottomrule		
    \end{tabularx}
    \label{tbl:intrinsic}
\end{table}

\subsection{Intrinsic Evaluation: Semantic AWEs}
\label{ssec:word_sim}

Table~\ref{tbl:intrinsic} presents the intrinsic scores of embeddings from the semantic AWE models, trained either from scratch (top section) or using multilingual transfer (bottom).
The benefit of multilingual transfer is evident in the scores of the projection and Cluster+Skipgram approaches, with the latter outperforming all other models regardless of the input features used or whether single or averaged embeddings are evaluated.
The single-sample performance $\rho_{\text{single}}$ is particularly
significant as it shows that individual representations can be compared accurately---a useful property for downstream applications such as semantic QbE~(\ref{ssec:sem_qbe}).

The ContrastiveRNN is the one exception that does not show a clear gain from initialising with multilingual weights compared to training from scratch.
As a sanity check, we evaluate the phonetic multilingual AWEs before semantic training (i.e.\ the foundation model used for transfer in the bottom section), obtaining a $\rho_{\text{single}} = \text{0.59\%}$ and $\rho_{\text{avg}} = -\text{0.13\%}$. 
As expected, this indicates that phonetic multilingual AWEs do not capture semantic information.
The table also shows the benefit of using self-supervised speech representations as input to AWEs instead of conventional features, as also in previous work~\cite{van_staden_comparison_2021, sanabria_analyzing_2023}; we use XSLR features from this point onwards.

Fig.~\ref{fig:sem_space} visualises the semantic embedding space of the Cluster+Skipgram model.
It is clear that the acoustic realisations of semantically related words end up in similar areas in the space. E.g.\ the model learned that spoken instances of ``orange", ``red", ``blue", ``yellow", and ``green"  should be close to each other.

\subsection{Extrinisic Evaluation: Semantic QbE}
\label{ssec:sem_qbe}
Table~\ref{tbl:qbe} compares the Cluster+Skipgram (semantic) and multilingual AWE (phonetic) models when used in a downstream QbE system.
We evaluate both exact and semantic QbE, where the latter gets awarded for retrieving exact query matches as well as utterances labelled as semantically related to the search query.
To situate results, we use a random baseline model that assigns a random relevance score to each utterance. (The relatively high scores of the random approach are due to the narrow domain of the evaluation data, Sec.~\ref{sec:setup}.) 

\begin{table}[!t]
	\mytable
	\caption{
        Exact and semantic QbE results (\%) on test data. 
        }
	\captionsep
  \begin{tabularx}{1\linewidth}{@{}lCCCCCCC@{}}

		\toprule				
		& \multicolumn{3}{c}{Exact QbE} & \multicolumn{4}{c}{Semantic QbE} \\
		\cmidrule(l){2-4} 
		\cmidrule(l){5-8}
		Model &  $P@10$ & $P@N$ & EER & $P@10$ & $P@N$ & EER & 
    $\rho$\\ 
		\midrule
				
		\underline{\textit{Baselines}:} \\
        Random & \hphantom{0}5.0 & \hphantom{0}5.0 &  50.0 &  \hphantom{0}9.1 & \hphantom{0}9.1 & 50.0 & \hphantom{0}5.7  \\

		Multilingual AWE & \textbf{90.7} & \textbf{82.8} & \textbf{\hphantom{0}4.7}  & \textbf{92.8} & \textbf{59.4} & 24.1 & 17.0  \\ [1mm]

            \underline{\textit{Semantic AWE model}:} \\
            Cluster+Skipgram & 85.8 & 71.3 & \hphantom{0}9.3  & 88.2 & 52.1 & \textbf{21.6} & \textbf{28.2} \\
            \bottomrule		
	\end{tabularx}
	\label{tbl:qbe}
\end{table}

\begin{table}[t]	
	\mytable
	\caption{Semantic QbE results (\%) on test data, where any instance of a query is masked from the search utterances.}
	\captionsep
 	\begin{tabularx}{\linewidth}{@{\extracolsep{5pt}}Lcccc}

		\toprule				
		Model & $P@10$ & $P@N$ & EER & Spearman's $\rho$ \\ 
		\midrule
				
		\underline{\textit{Baselines}:} & & & & \\
  		Random & \hphantom{0}9.2 & \hphantom{0}9.0 & 50.0 & \hphantom{0}5.4 \\
		Multilingual AWE & 21.5 & 15.6 & 46.1 & \hphantom{0}6.4\\[1mm]

            \underline{\textit{Semantic AWE model}:} \\
            Cluster+Skipgram & \textbf{29.9} & \textbf{23.1} & \textbf{32.3}  & \textbf{22.9}\\
  
            
            \bottomrule		
	\end{tabularx}
	\label{tbl:qbe_m}
\end{table}

Looking at the EER and Spearman's $\rho$ for semantic QbE, we see that the Cluster+Skipgram model achieves the highest score, outperforming the purely phonetic AWEs from the multilingual AWE model.
The reason why the phonetic multilingual AWE model outperforms the semantic model in $P@10$ and $P@N$ is due to its proficiency in detecting exact matches (which are also correct semantic matches).

To get a better sense of the ability of a model to retrieve non-verbatim semantic matches, we construct a difficult artificial semantic QbE task where we mask out all exact occurrences of the query word class in the search collection.
The results are shown in Table~\ref{tbl:qbe_m}.
Now we see a clear benefit in using the Cluster+Skipgram model, with the phonetic multilingual AWE model becoming close to random search. 

Our core goal was semantic QbE, but it is worth briefly touching on the exact QbE performance of the Cluster+Skipgram model in Table~\ref{tbl:qbe}. Although trained for semantics, this model still achieves reasonable exact retrieval performance, with only a drop of between 5\% and 10\% in scores compared to the multilingual AWE model. It is therefore clear that this semantic model is able to retain phonetic properties while also capturing semantic information related to context.

	\section{Conclusion}

We presented several semantic AWE modelling strategies. 
We specifically promoted transferring knowledge from a pre-trained multilingual AWE model trained for word-class discrimination.
Our best semantic AWE approach involves a soft clustering on the original multilingual AWEs, serving as input to a Skipgram-like model.
Through intrinsic and extrinsic evaluations, we demonstrated the effectiveness of our strategies in learning semantic representations from unlabelled speech data.
The main shortcoming of our work (as also in others~\cite{chung_speech2vec_2018}), is that the word segmentation is assumed to be known. 
This was reasonable given our goal of comparing different semantic AWE approaches on a sensible benchmark, but future work should look into incorporating unsupervised word segmentation methods~\cite{fuchs_unsupervised_2022, cuervo_contrastive_2022, kamper_word_2023} in order to do fully unsupervised semantic AWE modelling.

	\IEEEpeerreviewmaketitle

	\bibliographystyle{IEEEtran}
	\bibliography{spl2023}

\begin{thebibliography}{10}
\providecommand{\url}[1]{#1}
\csname url@samestyle\endcsname
\providecommand{\newblock}{\relax}
\providecommand{\bibinfo}[2]{#2}
\providecommand{\BIBentrySTDinterwordspacing}{\spaceskip=0pt\relax}
\providecommand{\BIBentryALTinterwordstretchfactor}{4}
\providecommand{\BIBentryALTinterwordspacing}{\spaceskip=\fontdimen2\font plus
\BIBentryALTinterwordstretchfactor\fontdimen3\font minus
  \fontdimen4\font\relax}
\providecommand{\BIBforeignlanguage}[2]{{%
\expandafter\ifx\csname l@#1\endcsname\relax
\typeout{** WARNING: IEEEtran.bst: No hyphenation pattern has been}%
\typeout{** loaded for the language `#1'. Using the pattern for}%
\typeout{** the default language instead.}%
\else
\language=\csname l@#1\endcsname
\fi
#2}}
\providecommand{\BIBdecl}{\relax}
\BIBdecl

\bibitem{mikolov_efficient_2013}
T.~Mikolov, K.~Chen, G.~Corrado, and J.~Dean, ``Efficient estimation of word
  representations in vector space,'' in \emph{Proc. {ICLR}}, 2013.

\bibitem{mikolov_distributed_2013}
T.~Mikolov, I.~Sutskever, K.~Chen, G.~Corrado, and J.~Dean, ``Distributed
  representations of words and phrases and their compositionality,'' in
  \emph{Proc. {NeurIPS}}, 2013.

\bibitem{pennington_glove_2014}
J.~Pennington, R.~Socher, and C.~Manning, ``{GloVe}: {Global} vectors for word
  representation,'' in \emph{Proc. {EMNLP}}, 2014.

\bibitem{sebastiani_machine_2002}
F.~Sebastiani, ``Machine learning in automated text categorization,'' \emph{ACM
  Computing Surveys}, vol.~34, pp. 1--47, 2002.

\bibitem{manning_introduction_2008}
C.~D. Manning, P.~Raghavan, and H.~Schütze, \emph{Introduction to information
  retrieval}.\hskip 1em plus 0.5em minus 0.4em\relax Cambridge University
  Press., 2008.

\bibitem{mikolov_exploiting_2013}
T.~Mikolov, Q.~V. Le, and I.~Sutskever, ``Exploiting similarities among
  languages for machine translation,'' \emph{arXiv preprint arXiv:1309.4168},
  2013.

\bibitem{conneau_word_2018}
A.~Conneau, G.~Lample, M.~Ranzato, L.~Denoyer, and H.~Jegou, ``Word translation
  without parallel data,'' in \emph{Proc. {ICLR}}, 2018.

\bibitem{levin_fixed-dimensional_2013}
K.~Levin, K.~Henry, A.~Jansen, and K.~Livescu, ``Fixed-dimensional acoustic
  embeddings of variable-length segments in low-resource settings,'' in
  \emph{Proc. {ASRU}}, 2013.

\bibitem{settle_discriminative_2016}
S.~Settle and K.~Livescu, ``Discriminative acoustic word embeddings:
  {Recurrent} neural network-based approaches,'' in \emph{Proc. {SLT}}, 2016.

\bibitem{chung_unsupervised_2016}
Y.-A. Chung, C.-C. Wu, C.-H. Shen, H.-Y. Lee, and L.-S. Lee, ``Unsupervised
  learning of audio segment representations using sequence-to-sequence
  autoencoder,'' in \emph{Proc. {Interspeech}}, 2016.

\bibitem{kamper_deep_2016}
H.~Kamper, W.~Wang, and K.~Livescu, ``Deep convolutional acoustic word
  embeddings using word-pair side information,'' in \emph{Proc. {ICASSP}},
  2016.

\bibitem{settle_query-by-example_2017}
S.~Settle, K.~Levin, H.~Kamper, and K.~Livescu, ``Query-by-example search with
  discriminative neural acoustic word embeddings,'' in \emph{Proc.
  {Interspeech}}, 2017.

\bibitem{kamper_truly_2019}
H.~Kamper, ``Truly unsupervised acoustic word embeddings using weak top-down
  constraints in encoder-decoder models,'' in \emph{Proc. {ICASSP}}, 2019.

\bibitem{jacobs_acoustic_2021}
C.~Jacobs, Y.~Matusevych, and H.~Kamper, ``Acoustic word embeddings for
  zero-resource languages using self-supervised contrastive learning and
  multilingual adaptation,'' in \emph{Proc. {SLT}}, 2021.

\bibitem{kamper_multilingual_2020}
H.~Kamper, Y.~Matusevych, and S.~Goldwater, ``Multilingual acoustic word
  embedding models for processing zero-resource languages,'' in \emph{Proc.
  {ICASSP}}, 2020.

\bibitem{hu_multilingual_2020}
Y.~Hu, S.~Settle, and K.~Livescu, ``Multilingual jointly trained acoustic and
  written word embeddings,'' in \emph{Proc. {Interspeech}}, 2020.

\bibitem{kamper_improved_2021}
H.~Kamper, Y.~Matusevych, and S.~Goldwater, ``Improved acoustic word embeddings
  for zero-resource languages using multilingual transfer,'' \emph{IEEE/ACM
  Trans. Audio, Speech, Lang. Process.}, vol.~29, pp. 1107--1118, 2021.

\bibitem{jacobs_multilingual_2021}
C.~Jacobs and H.~Kamper, ``Multilingual transfer of acoustic word embeddings
  improves when training on languages related to the target zero-resource
  language,'' in \emph{Proc. {Interspeech}}, 2021.

\bibitem{kamper_embedded_2017}
H.~Kamper, K.~Livescu, and S.~Goldwater, ``An embedded segmental {K}-means
  model for unsupervised segmentation and clustering of speech,'' in
  \emph{Proc. {ASRU}}, 2017.

\bibitem{jacobs_towards_2023}
C.~Jacobs, N.~C. Rakotonirina, E.~A. Chimoto, B.~A. Bassett, and H.~Kamper,
  ``Towards hate speech detection in low-resource languages: {Comparing} {ASR}
  to acoustic word embeddings on {Wolof} and {Swahili},'' in \emph{Proc.
  {Interspeech}}, 2023.

\bibitem{harwath_deep_2015}
D.~Harwath and J.~Glass, ``Deep multimodal semantic embeddings for speech and
  images,'' in \emph{Proc. {ASRU}}, 2015.

\bibitem{kamper_visually_2017}
H.~Kamper, S.~Settle, G.~Shakhnarovich, and K.~Livescu, ``Visually grounded
  learning of keyword prediction from untranscribed speech,'' in \emph{Proc.
  {Interspeech}}, 2017.

\bibitem{kamper_semantic_2019}
H.~Kamper, G.~Shakhnarovich, and K.~Livescu, ``Semantic speech retrieval with a
  visually grounded model of untranscribed speech,'' \emph{IEEE/ACM Trans.
  Audio, Speech, and Lang. Process.}, vol.~27, no.~1, pp. 89--98, 2019.

\bibitem{abdullah_integrating_2022}
B.~M. Abdullah, B.~Möbius, and D.~Klakow, ``Integrating form and meaning: {A}
  multi-task learning model for acoustic word embeddings,'' in \emph{Proc.
  {Interspeech}}, 2022.

\bibitem{chung_speech2vec_2018}
Y.-A. Chung and J.~Glass, ``{Speech2Vec}: {A} sequence-to-sequence framework
  for learning word embeddings from speech,'' in \emph{Proc. {Interspeech}},
  2018.

\bibitem{chen_phonetic-and-semantic_2019}
Y.-C. Chen, S.-F. Huang, C.-H. Shen, H.-y. Lee, and L.-s. Lee,
  ``Phonetic-and-semantic embedding of spoken words with applications in spoken
  content retrieval,'' in \emph{Proc. {SLT}}, 2019.

\bibitem{chen_simple_2020}
T.~Chen, S.~Kornblith, M.~Norouzi, and G.~Hinton, ``A simple framework for
  contrastive learning of visual representations,'' in \emph{Proc. {ICML}},
  2020.

\bibitem{chen_reality_2023}
G.~Chen and Y.~Cao, ``A reality check and a practical baseline for semantic
  speech embeddings,'' in \emph{{ICASSP}}, 2023.

\bibitem{rashtchian_collecting_2010}
C.~Rashtchian, P.~Young, M.~Hodosh, and J.~Hockenmaier, ``Collecting image
  annotations using {Amazon}'s {Mechanical} {Turk},'' in \emph{Proc. {NAACL}
  {HLT}}, 2010.

\bibitem{babu_xls-r_2022}
A.~Babu, C.~Wang, A.~Tjandra, K.~Lakhotia, Q.~Xu, N.~Goyal, K.~Singh, P.~von
  Platen, Y.~Saraf, J.~Pino, A.~Baevski, A.~Conneau, and M.~Auli, ``{XLS}-{R}:
  {Self}-supervised cross-lingual speech representation learning at scale,'' in
  \emph{Proc. {Interspeech}}, 2022.

\bibitem{van_staden_comparison_2021}
L.~van Staden and H.~Kamper, ``A comparison of self-supervised speech
  representations as input features for unsupervised acoustic word
  embeddings,'' in \emph{Proc. {SLT}}, 2021.

\bibitem{sanabria_analyzing_2023}
R.~Sanabria, H.~Tang, and S.~Goldwater, ``Analyzing acoustic word embeddings
  from pre-trained self-supervised speech models,'' in \emph{Proc. {ICASSP}},
  2023.

\bibitem{mcauliffe_montreal_2017}
M.~McAuliffe, M.~Socolof, S.~Mihuc, M.~Wagner, and M.~Sonderegger, ``Montreal
  forced aligner: {Trainable} text-speech alignment using {Kaldi},'' in
  \emph{Proc. {Interspeech}}, 2017.

\bibitem{ardila_common_2020}
R.~Ardila, M.~Branson, K.~Davis, M.~Henretty, M.~Kohler, J.~Meyer, R.~Morais,
  L.~Saunders, F.~M. Tyers, and G.~Weber, ``Common {Voice}: {A}
  massively-multilingual speech corpus,'' in \emph{Proc. {LREC}}, 2020.

\bibitem{agirre_study_2009}
E.~Agirre, E.~Alfonseca, K.~Hall, J.~Kravalova, M.~Paşca, and A.~Soroa, ``A
  study on similarity and relatedness using distributional and {WordNet}-based
  approaches,'' in \emph{Proc. {HLT}-{NAACL},}, 2009.

\bibitem{hill_simlex-999_2015}
F.~Hill, R.~Reichart, and A.~Korhonen, ``{SimLex}-999: {Evaluating} semantic
  models with (genuine) similarity estimation,'' \emph{Comput. Linguist.},
  vol.~41, no.~4, pp. 665--695, 2015.

\bibitem{fuchs_unsupervised_2022}
T.~S. Fuchs, Y.~Hoshen, and J.~Keshet, ``Unsupervised word segmentation using
  {K} nearest neighbors,'' in \emph{Proc. {Interspeech}}, 2022.

\bibitem{cuervo_contrastive_2022}
S.~Cuervo, M.~Grabias, J.~Chorowski, G.~Ciesielski, A.~Łańcucki,
  P.~Rychlikowski, and R.~Marxer, ``Contrastive prediction strategies for
  unsupervised segmentation and categorization of phonemes and words,'' in
  \emph{Proc. {ICASSP}}, 2022.

\bibitem{kamper_word_2023}
H.~Kamper, ``Word segmentation on discovered phone units with dynamic
  programming and self-supervised scoring,'' \emph{IEEE/ACM Trans. Audio,
  Speech, Lang. Process.}, vol.~31, pp. 684--694, 2023.

\end{thebibliography}
	
\end{document}